\documentclass[prd,aps,twocolumn,preprintnumbers,amsmath,amssymb,superscriptaddress]{revtex4}

\pdfoutput=1

\usepackage{amsmath}
\usepackage{graphicx}
\usepackage{dcolumn}
\usepackage{bm}
\usepackage{amssymb}
\usepackage{latexsym}
\usepackage{epstopdf}
\usepackage{threeparttable}
\usepackage{booktabs}
\usepackage{tabularx}
\usepackage{mathrsfs}
\usepackage{hhline}
\usepackage{multirow}
\usepackage{color}
\usepackage{lipsum}
\usepackage[colorlinks,linkcolor=magenta,anchorcolor=blue,citecolor=blue]{hyperref}

\def\blue{\color{blue}}

\def\be{\begin{equation}}
\def\ee{\end{equation}}
\def\bea{\begin{eqnarray}}
\def\eea{\end{eqnarray}}

\begin{document}
\title{Potential-driven Inflation with Disformal Coupling to Gravity}

\author{Taotao Qiu}
\email{qiutt@mail.ccnu.edu.cn}
\affiliation{Institute of Astrophysics, Central China Normal University, Wuhan 430079, China}

\author{Zehua Xiao}
\email{zehua2020@mails.ccnu.edu.cn}
\affiliation{Institute of Astrophysics, Central China Normal University, Wuhan 430079, China}

\author{Jiaming Shi}
\email{2016jimshi@mails.ccnu.edu.cn}
\affiliation{Institute of Astrophysics, Central China Normal University, Wuhan 430079, China}

\author{Muhsin Aljaf}
\email{mohsen@mail.ustc.edu.cn}
\affiliation{Department of Astronomy, University of Science and Technology of China, Hefei, Anhui 230026, China}

\begin{abstract}
In this paper, we investigate the potential-driven inflation models with a disformal coupling to Einstein Gravity, to find out the effects of such a coupling on these models. We consider a simple coupling form which introduces only one parameter, and three inflation models, namely the chaotic inflation, the Higgs inflation, and the monodromy inflation. We find that the disformal coupling can have some modifications to the observational variables of these models such as the power spectrum, the spectral index as well as the tensor/scalar ratio, although not too large due to the constraints on the disformal coupling parameter. With these modifications, one has the opportunity of improving models that lie on the edge of the favorable regions of Planck observational data, such as monodromy inflation. Moreover, the non-trivial sound speed of tensor perturbations (gravitational waves) may come out, due to the coupling of gravity and kinetic terms of the field. 
\end{abstract}

\maketitle
\section{introduction}
One of the most mysterious fields in our Universe might be the {\it inflaton} field \cite{Guth:1980zm, Albrecht:1982wi, Linde:1983gd} that played an important role in the early stages of cosmic evolution. By having a few assumptions such as slow-rolling, it can not only drive the fast acceleration (inflation) of the universe which simultaneously solves several Big-Bang problems, but also generate appropriate amount of primordial fluctuations that can explain CMB anisotropies and the formation of Large Scale Structures.

To extend its strength, inflation could also be coupled to gravity. The simplest case of such a consideration is the Brans-Dicke theory \cite{Brans:1961sx}, which later on has been extended to the so-called scalar-tensor theory \cite{Fujii:2003pa}, where what couples to gravity is an arbitrary function of the inflaton $\phi$. Such a case has been applied on many inflation models, the most famous of which might be that on Higgs inflation \cite{Bezrukov:2007ep}, where the inflaton field is recognized as the Higgs field appeared in Standard model of particle physics. In \cite{Bezrukov:2007ep} it was shown that such coupling could help Higgs inflation reconcile the inconsistency of constraints coming from particle physics and cosmology. Other forms of coupling also have been investigated, for instance, the nonminimal kinetic coupling where the kinetic term of the field couples to gravity, mainly via Einstein tensor, $G_{\mu\nu}$ \cite{Amendola:1993uh}, which can improve the model facing with the observational data \cite{Yang:2015pga}, and can also help the Higgs inflation alleviate the problem of unitarity \cite{Germani:2010gm} (also see applications to curvaton models \cite{Feng:2013pba}). More examples include coupling with Gauss-Bonnet term \cite{Lidsey:2003sj}, coupling to torsion \cite{Magueijo:2012ug, Cai:2011tc}, etc.  

Among these various couplings, we now consider a kind of ``disformal coupling", which is based on the so-called ``disformal" relationship between two metrics in terms of:
\be
\label{disformalmetric}
\tilde{g}_{\mu\nu}=A(\phi, X)g_{\mu\nu}+B(\phi,X)\partial_{\mu}\phi\partial_\nu\phi~,
\ee
where $\phi$ is some scalar field with its kinetic term $X\equiv-\partial_\mu\phi\partial^\mu\phi/2$, and $A$ and $B$ are arbitrary functions. This is a generalization of the ``conformal" relationship in which $B=0$, and is first proposed by Bekenstein \cite{Bekenstein:1992pj}. According to this relationship, when the gravity and matter parts belong to these two metrics separately, a coupling of the field to gravity/matter will appear as one discuss on either metric, which can cause physical significance. For instance, such a coupling is interestingly shown to cause variation of fundamental constants such as speed of light \cite{Clayton:1998hv, Bassett:2000wj, Magueijo:2003gj}, and may also lead to nontrivial sound speed of scalar perturbations and tensor perturbations in the early universe \cite{vandeBruck:2012vq, Creminelli:2014wna}. Therefore, it has also been applied to many aspects of cosmology, including inflation \cite{Kaloper:2003yf, vandeBruck:2015tna, Sato:2017qau, Karwan:2017vaw} (especially cosmological perturbation theory \cite{Minamitsuji:2014waa, Tsujikawa:2014uza, Motohashi:2015pra,Domenech:2015hka}), dark energy \cite{Koivisto:2008ak, Zumalacarregui:2010wj, Koivisto:2012za, Koivisto:2013fta, vandeBruck:2015ida, Bettoni:2015wla, Sakstein:2015jca, Mifsud:2017fsy}, dark matter \cite{Bettoni:2011fs, Bettoni:2012xv, Deruelle:2014zza, Hagala:2015paa, Gleyzes:2015pma}, massive gravity \cite{deRham:2010ik, Gannouji:2018aaw} and so on. In \cite{Bettoni:2013diz, Zumalacarregui:2013pma, Gleyzes:2014qga, DeFelice:2014bma, Bettoni:2015wta} the (in)variance under such a disformal transformation in Horndeksi and beyond Horndeski theories have been investigated, which help us better understand those general theories. See also \cite{vandeBruck:2013yxa, Brax:2013nsa, Brax:2014vva, Sakstein:2014isa, Brax:2015hma, Brax:2018bow, Dalang:2019fma} for various constraints on this coupling by astrophysical observations, and \cite{Sakstein:2015oqa} for recent reviews. 

In this paper, we consider the inflation models disformally coupled to gravity in a little bit more detail, namely how the astrophysical variables (such as power-spectrum $P_\zeta$, spectral index $n_s$ and tensor/scalar ratio $r$) of these models can be affected when such a coupling gets involved in. We consider only inflation which is driven by its potential in the form of power-law, with three typical examples: chaotic inflation, Higgs inflation and monodromy inflation. The rest of the paper is scheduled as follows: in Sec. II we performed the model in its general way and transform it to its Einstein frame, where we consider a simple form of coupling which contains only one more parameter, in Sec. III we take into account the explicit examples and calculated the observational variables such as power spectrum, spectral index, and tensor/scalar ratio. and show their relations with the disformal coupling parameter. We also discuss the effects on the sound speed of tensor perturbations. Sec. IV includes our conclusions.

\section{inflation with disformal coupling to gravity}
In this paper, we consider the inflation model with the following action:
\bea
\label{action}
S&=&\int d^{4}x\sqrt{-\tilde{g}}\left[\frac{m_{p}^{2}}{2}f(\phi)\tilde{R}-\frac{1}{2}\tilde{\nabla}_{\mu}\phi\tilde{\nabla}^{\mu}\phi-V(\phi)\right]\nonumber\\
&&+S[g_{\mu\nu}, \psi]~,
\eea
where $\tilde{g}$ and $g$ are the metric for gravity and matter part respectively, $\phi$ is viewed as inflaton field, and $\psi$ is the matter field. Here we also consider the nonminimal coupling of the inflaton field to Riemann scalar in terms of function $f(\phi)$, which will be useful in discussing some interesting inflation model such as Higgs inflation. If the two metrics are identical to each other, action (\ref{action}) will become that of a normal canonical inflation, which is quite trivial. However, when the two metrics are related by the disformal relationship as given in Eq. (\ref{disformalmetric}), nontrivial couplings will appear which makes the behavior of the field different.  

The difference in metric allows one to write the model in different frames. We denote the frame where the matter is minimally coupled with the field {\it the Jordan frame}, while that where the gravity is minimally coupled with the field  {\it the Einstein frame} \footnote{See \cite{Zumalacarregui:2012us} for the definition of different frames in disformal coupling theory.}. Different from the case of conformal coupling, in Jordan frame of disformal coupling models, gravity will be nontrivial coupled not only with inflaton field itself, but also with its derivative. In the following, we will consider how such a coupling will affect the behavior of inflation.
\subsection{background formulation}
For the sake of simplicity, we consider the functions $A$ and $B$ in relation (\ref{disformalmetric}) to be functions of $\phi$ only. Thus we furtherly have:   
\bea
\label{inversemetric}
\bar{g}^{\mu\nu}&=&\Omega^{-2}\left(g^{\mu\nu}-\frac{1}{1+\epsilon u^{2}}u^{\mu}u^{\nu}\right)~,\\
\sqrt{-\bar{g}}&=&\Omega^{4}\sqrt{1+\epsilon u^{2}}\sqrt{-g}~,
\eea
with $\Omega^{2}=A(\phi)$, $u^\mu\equiv \sqrt{B(\phi)/A(\phi)}\nabla^\mu\phi$, $u^\mu u_\mu=\epsilon u^2$. This can be useful in the reconstruction of action in the Jordan frame. 

It will be a tedious formulation to get the Jordan frame action, which we will leave in the appendix \ref{app}. However, as has been pointed in \cite{Sakstein:2015jca}, it can actually be embedded into the famous Horndeski action, where all the formulations for Horndeski theory are applicable. According to our calculation, the Horndeski-type Jordan frame action of our model reads:
\bea
\label{actionH}
S_{D}&=&\int d^{4}x\sqrt{-g}\Bigg\{K(\phi,X)+G_{3}(\phi,X)\Box\phi+G_{4}(\phi,X)R\nonumber\\
&&+G_{4,X}[(\Box\phi)^{2}-(\nabla_{\mu}\nabla_{\nu}\phi)(\nabla^{\mu}\nabla^{\nu}\phi)]\Bigg\}
\eea
with
\bea
K(\phi,X)&=&A(\phi)\Big\{\frac{[\omega(\phi)-m_p^2f(\phi)\mathcal{L}_{1}(\phi)]}{\sqrt{1-2g(\phi)^{2}X}}X+\nonumber\\
&&\sqrt{1-2g(\phi)^{2}X}[f(\phi)\mathcal{L}_{2}(\phi)X-V(\phi)A(\phi)]\Big\}~, \nonumber\\
G_3(\phi,X)&=&\frac{m_{p}^{2}}{2}\frac{\mathcal{L}_{3}(\phi)f(\phi)[-A(\phi)+4B(\phi)X]}{\sqrt{1-2g(\phi)^{2}X}}~, \nonumber  \\
G_4(\phi,X)&=&\frac{m_{p}^{2}}{2}A(\phi)f(\phi)\sqrt{1-2g(\phi)^{2}X}~,
\eea
here we also define the functions $\mathcal{L}_{i}(\phi),~i=1,2,3$ as:
\bea
\mathcal{L}_{1}(\phi)&\equiv&\ln A(\phi)\Big|_{\phi}^{2}+2\ln A(\phi)\Big|_{\phi}\ln f(\phi)\Big|_{\phi}+\frac{1}{2}\ln B(\phi)\Big|_{\phi}^{2}\nonumber\\
&&+\ln B(\phi)\Big|_{\phi}\ln f(\phi)\Big|_{\phi}~,\nonumber\\
\mathcal{L}_{2}(\phi)&\equiv&\ln[A(\phi)B(\phi)f(\phi)^{2}]\Bigg|_{\phi\phi}+\frac{1}{2}\ln[A(\phi)B(\phi)f(\phi)^{2}]\Bigg|_{\phi}^{2}~,\nonumber\\
\mathcal{L}_{3}(\phi)&\equiv&\ln[A(\phi)B(\phi)f(\phi)^{2}]\Bigg|_{\phi}~,
\eea
where $|_{\phi}$ denotes derivative with respect to $\phi$.

This is a quite complicated action, and as one can see, the kinetic term appeared in $G_4$, so it will be different from normal nonminimal coupling inflation models. However, we set out a simple case of disformal transformation (\ref{disformalmetric}), by assuming $A(\phi)=f(\phi)^{-1}$, and $B(\phi)/A(\phi)=M=const.$. The choice of $A$ is to compensate for the nonminimal coupling in the original action (\ref{action}), as was done for Higgs inflation models \cite{Bezrukov:2007ep}. Therefore the transformation brings only one parameter, $M$, and this is one of the minimal extensions of conformal transformation where $M=0$. Note that $M$ actually stands for the scale of interaction between the kinetic term of the field and gravity, or in other words, the scale where the nontrivial part of disformal coupling takes part. In such a case, the Jordan frame action will be Eq. (\ref{actionH}) with
\bea
\label{functions}
K(\phi,X)&=&\frac{\bar{K}(\phi)X}{\sqrt{1-2MX}}-\bar{V}(\phi)\sqrt{1-2MX}~,\nonumber \\
G_{3}(\phi,X)&=&0~,\nonumber\\
G_{4}(\phi,X)&=&\frac{m_{p}^{2}}{2}\sqrt{1-2MX}~,
\eea
with ${\cal L}_1=-(3/2)[\ln f(\phi)]|_\phi^2$, ${\cal L}_2={\cal L}_3=0$. Moreover, we defined 
\be
\bar{K}(\phi)=\frac{\omega(\phi)}{f(\phi)}+\frac{3m_{p}^{2}}{2}\frac{f_{\phi}(\phi)^{2}}{f(\phi)^{2}}~,~\bar{V}(\phi)=\frac{V(\phi)}{f(\phi)^{2}}~.
\ee

One can get the Friedmann equations and equation of motion of this model straightforwardly from such an action, as the most general case has been discussed in \cite{Kobayashi:2011nu, DeFelice:2011uc}. According to Eqs. (\ref{functions}), the Friedmann equations:
\bea
\label{friedmann1}
3H^{2}&=&\frac{4H^{2}G_{4}\delta_{KX}-K}{2G_{4}(1-8\delta_{G4X}-8\delta_{G4XX})}~,\nonumber\\
&=&\frac{\bar{K}X+(1-2MX)\bar{V}}{m_{p}^{2}}~,\\
\label{friedmann2}
-(3H^{2}+2\dot{H})&=&\frac{1}{2G_{4}(1-4\delta_{G4X})}\Big[K+4H^{2}G_{4}(-4\delta_{G4X}\delta_{\phi}\nonumber\\
&&-8\delta_{G4XX}\delta_{\phi})\Big]\nonumber\\
&=&\frac{\bar{K}X-(1-2MX)\bar{V}}{m_{p}^{2}}-8H^{2}\delta_{G4X}\delta_{\phi}~,
\eea
and the equation of motion:
\be
\dot{\cal J}+3H{\cal J}-{\cal P}=0~,
\ee
where
\bea
{\cal J}&=&\dot{\phi}K_{X}+6H^{2}\dot{\phi}(G_{4X}+2XG_{4XX})=\frac{\bar{K}\dot{\phi}}{(1-2MX)^{1/2}}~,\\
{\cal P}&=&K_{\phi}=\frac{\bar{K}_\phi X}{\sqrt{1-2MX}}-\bar{V}_\phi\sqrt{1-2MX}~.
\eea
Note that the equation of motion is also equivalent to the second-order equation form: $\ddot{\phi}+(1-2MX)[3H\dot{\phi}+(1-2MX)(\bar{V}_{\phi}/\bar{K})+X\bar{K}_{\phi}/\bar{K}]=0$.

\subsection{perturbations}
From action (\ref{actionH}), it is also straightforward to get the scalar and tensor perturbations. The perturbed line element with metric $g_{\mu\nu}$ is written as:
\be
ds^2=-N^2dt^2+a^2(t)e^{2\zeta\delta_{ij}+2\gamma_{ij}}(dx^i+N^idt)(dx^j+N^jdt)~
\ee
where $N$ and $N^i$ are the lapse function and shift vector, while $\zeta$ and $\gamma_{ij}$ are scalar and tensor perturbations, respectively. Moreover, it is useful to define the so-called ``slow-varying" parameters \cite{DeFelice:2011uc}: 
\bea
\label{slowvarying}
&&\delta_{\phi}\equiv\frac{\ddot{\phi}}{H\dot{\phi}}~,\delta_{G4}\equiv\frac{\dot{G}_4}{HG_4}~,\nonumber\\
&&\delta_{KX}\equiv\frac{K_{X}X}{2H^{2}G_{4}}=\frac{X[(1-MX)\bar{K}+M(1-2MX)\bar{V}]}{m_{p}^{2}H^{2}(1-2MX)^{2}}~,\nonumber\\
&&\delta_{KXX}\equiv\frac{K_{XX}X^{2}}{2H^{2}G_{4}}\nonumber\\
&&=\frac{MX^{2}[(2-MX)\bar{K}+M(1-2MX)\bar{V}]}{m_{p}^{2}H^{2}(1-2MX)^{3}}~,\nonumber\\
&&\delta_{G4X}\equiv\frac{G_{4X}X}{2G_{4}}=-\frac{MX}{2(1-2MX)}~,\nonumber\\
&&\delta_{G4XX}\equiv\frac{G_{4XX}X^{2}}{2G_{4}}=-\frac{M^{2}X^{2}}{2(1-2MX)^{2}}~,\nonumber\\
&&\delta_{G4XXX}\equiv\frac{G_{4XXX}X^{3}}{2G_{4}}=-\frac{3M^{3}X^{3}}{2(1-2MX)^{3}}~,
\eea
for later use. We see that there are some simple but useful relations between those parameters, namely
\be
\delta_{G4}=4\delta_{G4X}\delta_{\phi}~,~~~\delta_{G4XX}=-2\delta_{G4X}^{2}~,~~~\delta_{G4XXX}=12\delta_{G4X}^{3}~.
\ee

The perturbed action for tensor perturbation is \cite{Kobayashi:2011nu,DeFelice:2011uc}:
\be
\label{tensorpert}
S_{T}^{(2)}=\frac{1}{8}\int dtd^{3}xa^{3}[{\cal G}_{T}\dot{\gamma}_{ij}^{2}-a^{-2}{\cal F}_{T}(\nabla\gamma_{ij})^{2}]
\ee
where in our model,  
\bea
{\cal G}_{T}&=&\frac{m_{p}^{2}}{\sqrt{1-2MX}}~,\\
{\cal F}_{T}&=&m_{p}^{2}\sqrt{1-2MX}~.
\eea
The sound speed squared for tensor perturbation is
\be
c_{T}^{2}\equiv\frac{{\cal F}_{T}}{{\cal G}_{T}}=1-2MX~,
\ee
which means that unlike the conformal coupling models, in the disformal coupling models the sound speed of tensor mode will deviate from 1, which indicates a different propagation speed of gravitational waves than that of light. The equation of motion for $\gamma_{ij}$ derived from action (\ref{tensorpert}) is:
\be
\gamma_{ij}^{\prime\prime}-c_T^2\nabla^2\gamma_{ij}+\frac{(a^2{\cal G}_T)^\prime}{a^2{\cal G}_T}\gamma_{ij}^\prime=0~,
\ee
with the solution
\be
\gamma_{ij}=\text{constant}~,~~~\int\frac{dt}{a^3(t){\cal G}_T}~.
\ee
Therefore the power spectrum for tensor perturbations is
\bea
\label{tensorspectrum}
P_{T}&\equiv&\frac{k^3}{2\pi^2}|\gamma_{ij}|^2=\frac{2H^{2}}{{\cal F}_Tc_T\pi^{2}}\nonumber\\
	&=&\frac{2H^{2}}{\pi^{2}m_{p}^{2}(1-2MX)}~,
\eea
and the spectral tilt:
\be
n_{T}\equiv\frac{d\ln P_T}{d\ln k}=-2\epsilon+\frac{1}{2}g_{T}-\frac{3}{2}f_{T}~,
\ee
where $f_{T}\equiv\dot{\cal F}_{T}/(H{\cal F}_{T})$ and $g_{T}\equiv\dot{\cal G}_{T}/(H{\cal G}_{T})$ are defined.

The scalar perturbation can be obtained in like manner. The perturbed action for scalar perturbation is \cite{Kobayashi:2011nu,DeFelice:2011uc}:
\be
\label{scalarpert}
S_{S}^{(2)}=\int dtd^{3}xa^{3}[{\cal G}_{S}\dot{\zeta}^{2}-a^{-2}{\cal F}_{S}(\nabla\zeta)^{2}]
\ee
where in our model, 
\bea
\label{GS}
{\cal G}_{S}&=&2G_{4}\left[6\delta_{G4X}+\frac{\delta_{KX}+2\delta_{KXX}}{(1-4\delta_{G4X})^{2}}\right]~\nonumber\\
&=&2G_{4}\frac{\bar{K}X}{m_{p}^{2}H^{2}(1-2MX)}~,\\
\label{FS}
{\cal F}_{S}&=&2G_{4}(\epsilon+4\delta_{G4X}\delta_{\phi})=2G_{4}\frac{\bar{K}X}{m_{p}^{2}H^{2}}~,
\eea
where in the last steps of Eqs. (\ref{GS}) and (\ref{FS}) we've made use of the definitions of the slow-varying parameters in (\ref{slowvarying}), as well as the Friedmann equations (\ref{friedmann1})-(\ref{friedmann2}), from which we can get:
\be
\label{KX}
\bar{K}X=H^{2}m_{p}^{2}(\epsilon+4\delta_{G4X}\delta_{\phi})~.
\ee

The sound speed for scalar perturbation is therefore
\be
c_{S}^{2}\equiv\frac{{\cal F}_{S}}{{\cal G}_{S}}=1-2MX~,
\ee
which has the same behavior as that of $c_T^2$. Moreover, similar to the tensor perturbation, the equation of motion for $\zeta$ derived from action (\ref{scalarpert}) is:
\be
\zeta^{\prime\prime}-c_S^2\nabla^2\zeta+\frac{(a^2{\cal G}_S)^\prime}{a^2{\cal G}_S}\zeta^\prime=0~,
\ee
with the solution
\be
\zeta=\text{constant}~,~~~\int\frac{dt}{a^3(t){\cal G}_S}~.
\ee
Therefore the power spectrum for scalar perturbations is
\bea
\label{scalarspectrum}
P_{S}&\equiv&\frac{k^3}{2\pi^2}|\gamma_{ij}|^2=\frac{H^{2}}{8{\cal F}_Sc_S\pi^{2}}\nonumber\\
&=&\frac{H^{4}}{8\pi^{2}\bar{K}X(1-2MX)}~,
\eea
and the spectral tilt:
\bea
\label{index}
n_{S}&\equiv&1+\frac{d\ln P_S}{d\ln k}=1-2\epsilon+\frac{1}{2}g_{S}-\frac{3}{2}f_{S}~\nonumber\\
&=&1-\frac{4\bar{K}X}{H^2m_p^2}-\frac{\dot{\bar{K}}}{H\bar{K}}-\frac{2}{1-2MX}\delta_\phi~,
\eea
where $f_{S}\equiv\dot{\cal F}_{S}/(H{\cal F}_{S})$ and $g_{S}\equiv\dot{\cal G}_{S}/(H{\cal G}_{S})$ are defined.

Moreover, from (\ref{tensorspectrum}) and (\ref{scalarspectrum}) we can get the tensor/scalar ratio:
\be
\label{tsratio}
r\equiv\frac{P_T}{P_S}=16\frac{\bar{K}X}{m_{p}^{2}H^{2}}~.
\ee
Note that when $G_4=m_p^2/2$, $\delta_{G4X}$ will be vanishing in Eq. (\ref{KX}), and thus $r=16\epsilon$ which recovers the consistency relation.

\section{applying to potential driven models}
\subsection{models with two ``slow-roll conditions"}
In this section, we consider the application of the above analysis to a large category of inflation models, which is driven mainly by its potential. Usually, in inflation models (especially slow-roll models), two kinds of ``slow-roll conditions" will be admitted: one is to let $\bar{K}X=\bar{K}\dot\phi^2/2\ll \bar{V}$. In this case, the first Friedmann equations (\ref{friedmann1}) can be rewritten as
\be
\label{friedmannsr1}
3m_p^2H^2\simeq (1-2MX)\bar{V}~.
\ee
The other is to let $\ddot\phi\ll H\dot\phi$. In this case, the second Friedmann equation (\ref{friedmann2}) and the equation of motion now becomes: 
\bea
\label{friedmannsr2}
&&\dot H\simeq-\bar{K}X/m_p^2~,\\
&&(1-2MX)\left[3H\dot\phi+(1-2MX)\frac{\bar{V}_\phi}{\bar{K}}\right]\simeq 0~,
\eea
from which we get 
\be
\label{dotphi}
\dot\phi\simeq -(1-2MX)\frac{\bar{V}_\phi}{3H\bar{K}}~.
\ee

If the two slow-roll conditions are both satisfied, like most cases of inflation models, from the above we can get the slow-roll parameter
\be
\epsilon\equiv-\frac{\dot H}{H^2}\simeq \frac{m_p^2}{2\bar{K}}\frac{\bar{V}_\phi^2}{\bar{V}^2}~,
\ee
and for minimal coupling case in the original action (\ref{action}) where $\bar{K}=1$, it reduces to the normal slow-roll inflation case. The efolding-number which describes the duration of inflation now reads:
\be
N\equiv\int_{t_e}^{t_0} Hdt=\int_{\phi_e}^{\phi_0} \frac{H}{\dot\phi}d\phi=\frac{1}{m_p^2}\int_{\phi_e}^{\phi_0} \frac{\bar{V}}{\bar{V}_\phi}\bar{K}d\phi~,
\ee
where the subscript $0$ and $e$ represent the begining and ending time of inflation, respectively. Since at the end of inflation we have $\epsilon(\phi_e)=1$, and by requiring $N(\phi_0, \phi_e)=60$ we can get $\phi_0$, the initial value of $\phi$ where inflation starts. This is a very standard way of getting initial conditions that satisfy slow-roll conditions.

Note that for both two quantities, the factor $1-2MX$ in (\ref{friedmannsr1}) and (\ref{dotphi}) compensate to each other, and actually do not have any effect. Moreover, from the results in last section one can get:
\be
P_{S}=\frac{\bar{K}\bar{V}^{3}}{12\pi^{2}m_{p}^{6}\bar{V}_{\phi}^{2}}~,~n_s=1-2\frac{m_p^2\bar{V}_{\phi}^{2}}{\bar{V}^{2}\bar{K}}~,
~r=\frac{8m_p^2\bar{V}_{\phi}^{2}}{\bar{V}^{2}\bar{K}}~,
\ee
namely the observable quantities such as scalar and tensor power spectrum does not depend on the parameter of disformal coupling $M$ either. Although the dependence appears in form of $1-2MX$ in Eqs. (\ref{scalarspectrum}), (\ref{index}) and (\ref{tsratio}), it is again compensated by the dependence of the initial conditions, namely $\dot\phi$ (or $X$) in (\ref{dotphi}). Therefore in the case where both the two ``slow-roll conditions" are rigidly satisfied, the disformal coupling could not actually make any difference on the observables, although it does have effects on sound speeds of both tensor and scalar perturbations. This is also in consistency with the conclusion made in \cite{Minamitsuji:2014waa, Tsujikawa:2014uza, Motohashi:2015pra,Domenech:2015hka} that the perturbations are invariant under such a disformal transformation. 

As a side remark, using the Friedmann equation (\ref{friedmannsr1}), one can get another form of $\dot\phi$:
\be
\dot\phi\simeq -\frac{m_p\bar{V}_\phi}{\sqrt{3\bar{V}}\bar{K}}\left(1+M\frac{m_p^2\bar{V}_\phi^2}{3\bar{V}\bar{K}^2}\right)^{-\frac{1}{2}}~,
\ee
where $\dot\phi_0=\dot\phi(\phi_0)$ gives initial condition for $\dot\phi$. From the expression we can see that, to guarantee that $\dot\phi$ is a real number, the paramater $M$ should satisfy 
\be
0\geq M\geq-\frac{3\bar{V}\bar{K}^2}{\bar{V}_\phi^2}~,
\ee
Moreover, for $|M|\ll 1$, the initial value of $\dot\phi$ reduces to that of standard slow-roll condition as expected, for $|M|\gg 1$, on the other hand, we have $\dot\phi=-1/\sqrt{M}$, nearly a constant. 

\subsection{models with one ``slow-roll condition" violated}
Since we are considering potential-driven inflation, we identify that the first ``slow-roll condition" should be satisfied. However, there will sometimes be cases that the second condition is not satisfied, for instance, in ultra-slow-roll/constant-roll inflation models \cite{Kinney:2005vj}, $\ddot\phi$ and $H\dot\phi$ will be of the same order. In those cases, we cannot get the initial condition of $\dot\phi$ by relying on Eq. (\ref{dotphi}). Another choice of initial condition is to get a somehow ``fixed" value of ($\phi_0$, $\dot\phi_0$), by setting them to be the ones calculated when, say, $M=0$. By such an imposing we screened the effects of $M$ on the initial values, which can give us pure $M$-dependence on observable quantities during field evolution. 

\subsubsection{chaotic inflation}
We will first consider the chaotic inflation, proposed by A. Linde in the 1980's \cite{Linde:1983gd}, which is among the earliest inflation models. The model requires
\be
f(\phi)=1~,~~~V(\phi)=\frac{\lambda}{4}\phi^4~,
\ee
which makes it easy to find out the initial values of $\phi$ and $\dot\phi$:
\be
\label{initialchaotic}
\phi_{0}=2\sqrt{2(N+1)}m_{p}~,~~~\dot{\phi}_{0}=-4\sqrt{\frac{2\lambda(N+1)}{3}}m_{p}^{2}~,
\ee
where $N$ is the efolding number of the inflation. The current observation is normalized at the time point when $N=60$. Moreover, applying the first slow-roll condition (\ref{friedmannsr1}) to Eqs. (\ref{scalarspectrum}), (\ref{index}) and (\ref{tsratio}), the scalar spectrum, spectral index and the tensor/scalar ratio are
\bea
P_{S}&=&\frac{32(1-2MX)\lambda^{2}(N+1)^{4}m_{p}^{4}}{9\pi^{2}X}~,\\
n_{S}&=&7-6\sqrt{1-2MX}\nonumber\\
&&-\frac{3X}{4\lambda(1-2MX)(N+1)^{2}m_{p}^{4}}~,\\
r&=&\frac{3X}{(1-2MX)\lambda(N+1)^{2}m_{p}^{4}}~,
\eea
From which one can see the $M$-dependence of the perturbation variables. Since $M$ is a negative value, one can see that when $|M|$ gets enlarged, it will cause an enhancement of $P_S$ (scalar spectrum) and a suppression of $r$ (tensor/scalar ratio) or vice versa, while $P_T$ (tensor spectrum) remains almost unchanged with regard to $M$. For enough large $|M|$ so as that $|2MX|\gg 1$, one has 
\bea
P_{S}&\simeq&\frac{64|M|\lambda^{2}(N+1)^{4}m_{p}^{4}}{9\pi^{2}}~,\\
n_{S}&\simeq&-8\sqrt{6|M|\lambda(N+1)}m_{p}^{2}~,\\
r&\simeq&\frac{3}{2|M|\lambda(N+1)^{2}m_{p}^{4}}~,
\eea
where one can see more clearly the relationship of $P_S\propto|M|$, $n_{S}\propto\sqrt{|M|}$, $r\propto|M|^{-1}$. Moreover, for large $|M|$, $\lambda$ is expected to get smaller value in order to meet with the observational data.

\begin{widetext}
\begin{figure*}
\centering
\includegraphics[scale=0.55]{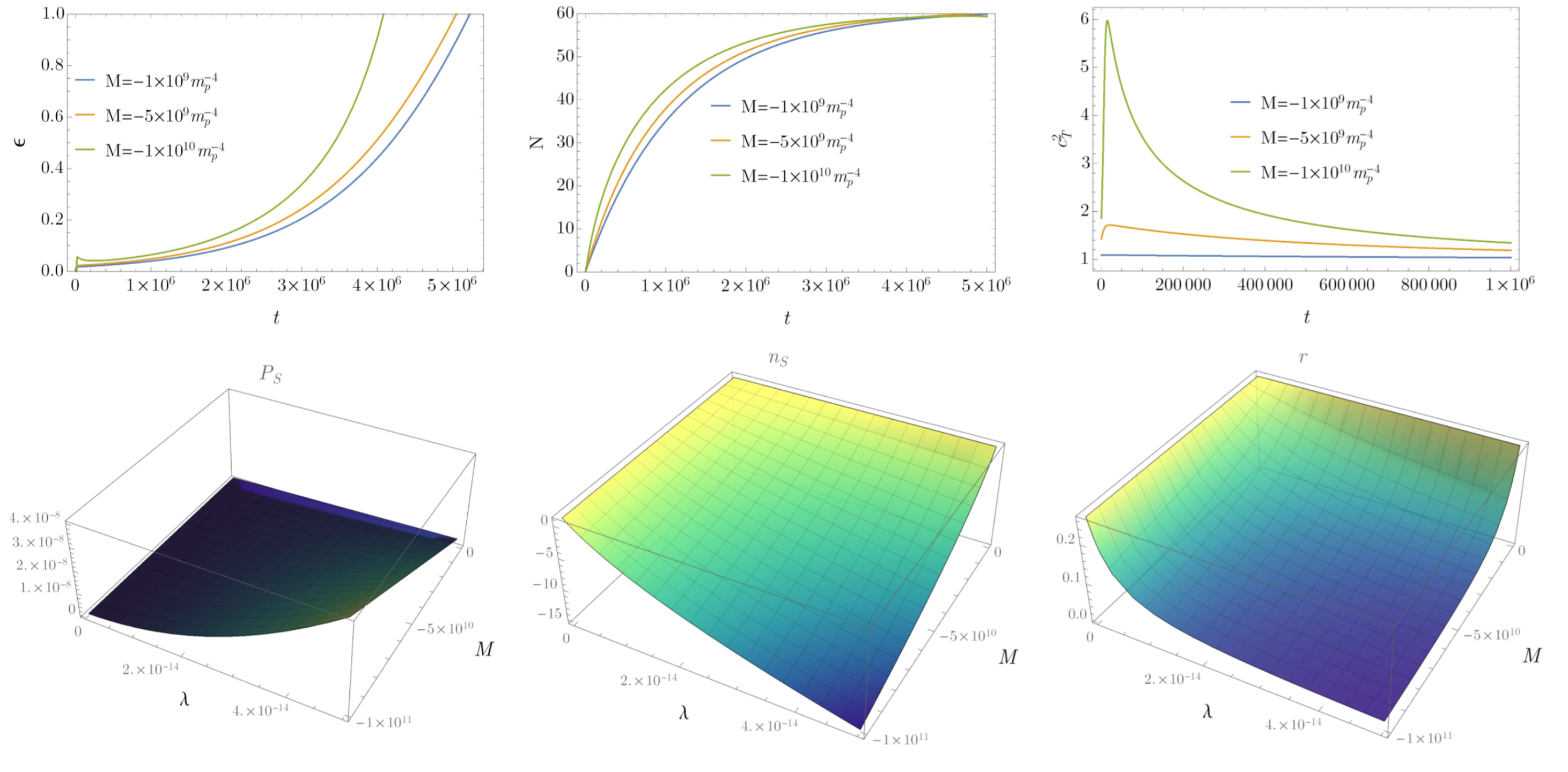}
\caption{Upper panel: The evolution of the slow-roll parameter $\epsilon$, the e-folding number $N$ as well as the sound speed squared of tensor perturbation $c_T^2$ of chaotic inflation model, with $\phi_0\simeq 22m_p$, $\dot\phi_0\simeq -9.26\times 10^{-6}m_p^2$, and blue, orange and green lines correspond to $M=-10^{9}m_p^{-4}$, $M=-5\times 10^{9}m_p^{-4}$, $M=-10^{10}m_p^{-4}$, respectively. Down panel: The scalar power spectrum $P_S$, its spectral index $n_S$ and tensor/scalar ratio $r$ as functions of the parameters $\lambda$ and $M$.  }\label{plotchaotic}
\end{figure*}
\end{widetext}

In Fig. \ref{plotchaotic}, we plot the evolutions of various background variables such as slow-roll parameter $\epsilon$, the e-folding number $N$ and the sound speed squared of tensor perturbation $c_T^2$ of this model, with initial conditions chosen according to (\ref{initialchaotic}), considering three cases of $M$ being $-10^{9}m_p^{-4}$, $-5\times 10^{9}m_p^{-4}$, and $-10^{10}m_p^{-4}$. From the plot we can see that, although different in $M$, all the three cases can have inflation for $N$ up to $60$, while as $|M|$ grows, $c_T^2$ will deviate from unity. We also plot the perturbation variables such as the scalar power spectrum $P_S$, its spectral index $n_S$ and the tensor/scalar ratio $r$ in terms of $\lambda$ and $M$, and we can see their dependence on $\lambda$ and $M$ are consistent with the above analytical analysis.

\begin{figure}
\centering
\includegraphics[scale=0.3]{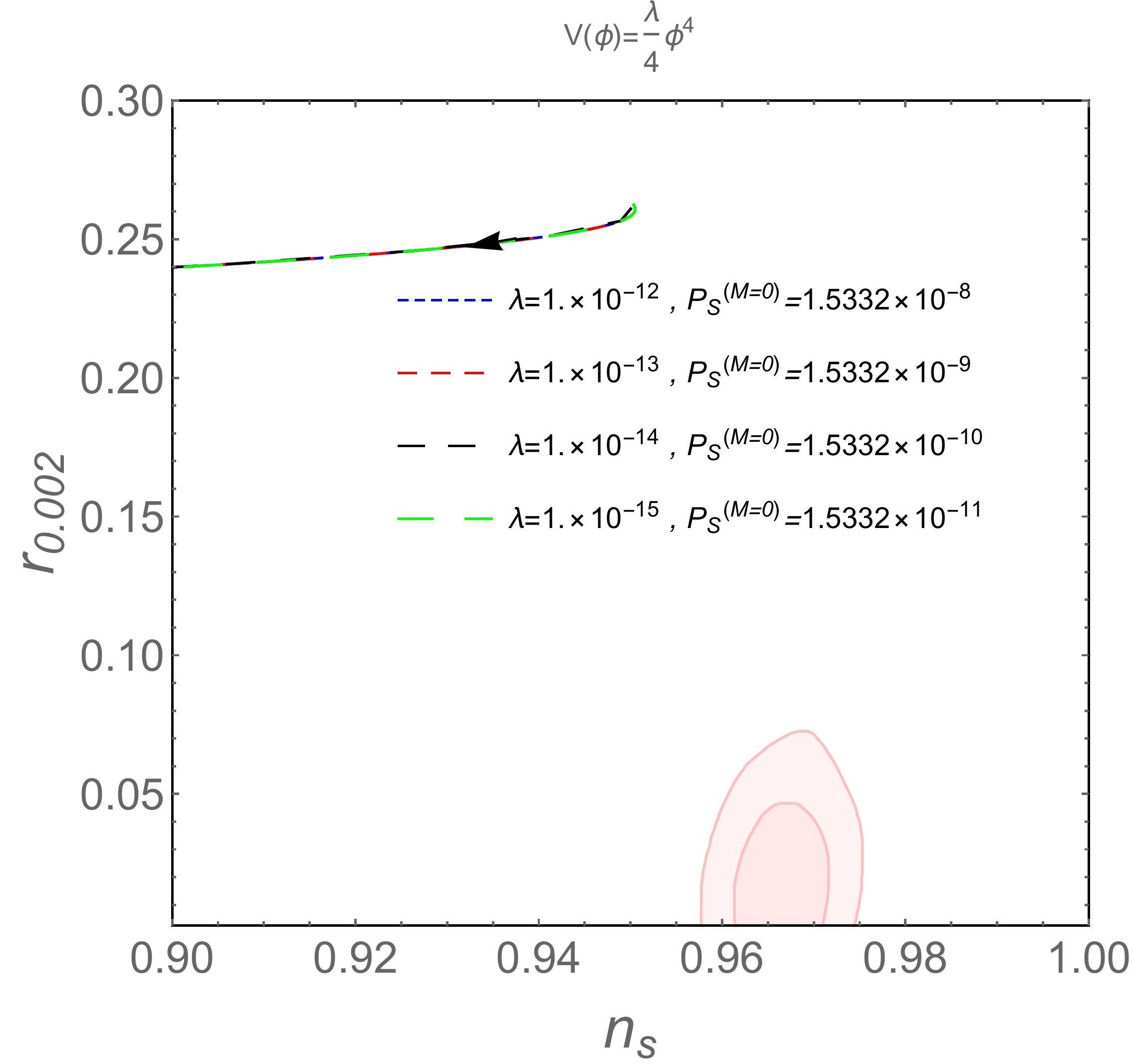}
\caption{$n_s$-$r$ constraints on chaotic inflation. The black arrows represent the direction of $|M|$ growth. The pink region represents TT,TE,EE+lowE+lensing+BK15+BAO data constraints (68\% and 95\% CL) for $n_s$ and $r$ from 2018 Planck Collaboration data \cite{Akrami:2018odb}. Here we choose $N=60$, while $\lambda$ is set as $10^{-12}$, $10^{-13}$, $10^{-14}$ and $10^{-15}$, respectively. $P_S^{(M=0)}$ means the value of $P_S$ when $M=0$.}\label{Cinf}
\end{figure}

In Fig. \ref{Cinf} we compare $n_S$ and $r$ in this model with the newest Planck observational data. We show that as $|M|$ grows, both $n_S$ and $r$ decrease (as also shown in the last plot), but $n_S$ has a faster speed. Although the decrease of $r$ tends to alleviate the dilemma of having too large $r$ for chaotic inflation models, it may not help much if one also wants to keep $n_S$ inside the allowable region. Maybe a better way is to have a more complicated form of disformal coupling with more parameters involved, however, this goes beyond the scope of the current study and can be postponed in the future.

\subsubsection{Higgs inflation}
We then consider the Higgs inflation, where it is proposed that inflaton is acted by the Higgs boson. Although the potential form is the same as that of chaotic inflation, due to the constraints from particle physics we have $\lambda\simeq 0.13$, which cannot guarantee the model be consistent with the constraints from cosmological observations. As has been suggested in \cite{Bezrukov:2007ep}, this problem can be circumvented by introducing a nonminal coupling term, namely in this model we have 
\be
f(\phi)=1+\xi\left(\frac{\phi}{m_p}\right)^2~,~~~V(\phi)=\frac{\lambda}{4}\phi^4~.
\ee
Due to the involvement of $f(\phi)$, the analytical solution of $\phi_0$ and $\dot\phi_0$ becomes complicated. However, since the usual case we can apply the approximation of $\xi\gg 1$, from which we get
\be
\label{initialhiggs}
\phi_{0}\simeq\sqrt{\frac{2(2N+\sqrt{3})}{3\xi}}m_{p}~,~\dot{\phi}_{0}\simeq-\sqrt{\frac{\lambda}{18(2N+\sqrt{3})\xi^{3}}}m_{p}^{2}~,
\ee
while the scalar spectrum, spectral index and the tensor/scalar ratio are:
\bea
\label{pshiggs}
P_{S}&\simeq&\frac{(1-2MX)\lambda^{2}(2N+\sqrt{3})^{3}m_{p}^{4}}{2592\pi^{2}X\xi^{5}[3+2(2N+\sqrt{3})]^{2}}~,\\
\label{nshiggs}
n_{S}&\simeq&7-6\sqrt{1-2MX}\nonumber\\
&&+\frac{432\xi^{3}X}{(2N+\sqrt{3})(1-2MX)\lambda m_{p}^{4}}~,\\
\label{rhiggs}
r&\simeq&\frac{1728X\xi^{3}}{(1-2MX)\lambda(2N+\sqrt{3})m_{p}^{4}}~,
\eea
where the $M$-dependence of the perturbation variables are the same as the first case, namely will not be changed by the involvement of nonminimal coupling. This can also be seen from the approximations for $|2MX|\gg 1$, where
\bea
P_{S}&\simeq&\frac{|M|\lambda^{2}(2N+\sqrt{3})^{3}m_{p}^{4}}{1296\pi^{2}\xi^{5}[3+2(2N+\sqrt{3})]^{2}}~,\\
n_{S}&\simeq&-\sqrt{\frac{2|M|m_{p}^{4}\lambda}{(2N+\sqrt{3})\xi^{3}}}~,\\
r&\simeq&\frac{864\xi^{3}}{|M|\lambda(2N+\sqrt{3})m_{p}^{4}}~.
\eea

On the other hand, it is known that in Higgs inflation, although $\lambda$ is nearly fixed by the particle physics, the value of $\xi$ could be large enough to change the results. One can also obtain the approximations for large $\xi$, namely $\xi\gg 1$, from (\ref{pshiggs})-(\ref{rhiggs}), which are:
 \bea
P_{S}&\simeq&\frac{\lambda(2N+\sqrt{3})^{4}}{72\pi^{2}\xi^{2}[3+2(2N+\sqrt{3})]^{2}}~,\\
n_{S}&\simeq&1-\frac{12}{(2N+\sqrt{3})^{2}}~,\\
r&\simeq&\frac{48}{(2N+\sqrt{3})^{2}}~,
\eea
where all the $M$ dependence disappears. The reason might be that for large $\xi$, the potential is so flattened to be closed to a constant, and any effects on the perturbations are dissipated away. 

\begin{widetext}
\begin{figure*}
\centering
\includegraphics[scale=0.55]{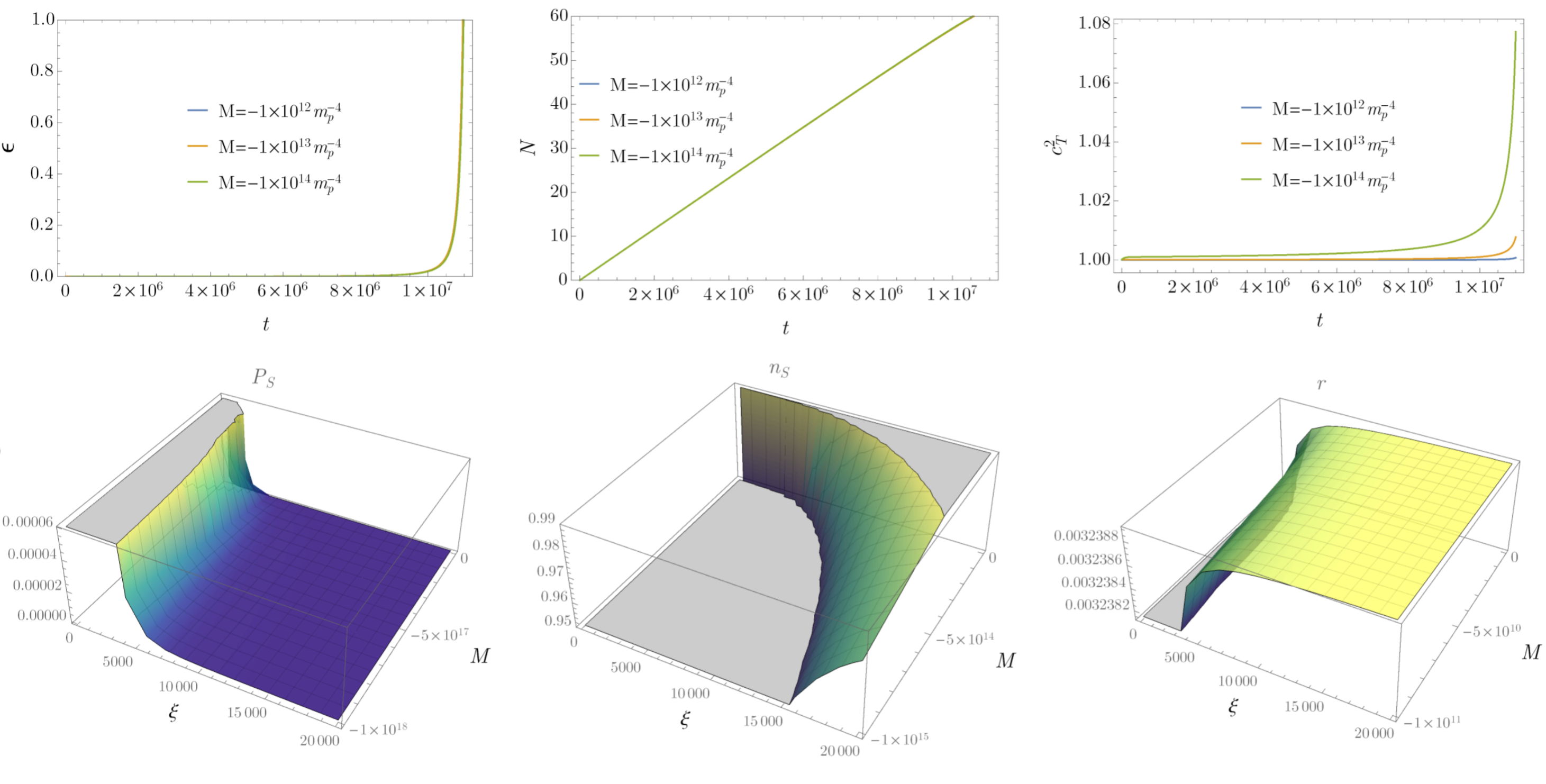}
\caption{Upper panel: The evolution of the slow-roll parameter $\epsilon$, the e-folding number $N$ as well as the sound speed squared of tensor perturbation $c_T^2$ of Higgs inflation model, with $\phi_0\simeq 0.07m_p$, $\dot\phi_0\simeq -3.21\times 10^{-9}m_p^2$, and blue, orange and green lines correspond to $M=-10^{12}m_p^{-4}$, $M=-10^{13}m_p^{-4}$, $M=-10^{14}m_p^{-4}$, respectively. Down panel: The scalar power spectrum $P_S$, its spectral index $n_S$ and tensor/scalar ratio $r$ as functions of the parameters $\xi$ and $M$. }\label{plothiggs}
\end{figure*}
\end{widetext}

In Fig. \ref{plothiggs}, we plot the evolutions of slow-roll parameter $\epsilon$, the e-folding number $N$ and the sound speed squared of tensor perturbation $c_T^2$ of this model, with initial conditions chosen according to (\ref{initialhiggs}), considering three cases of $M$ being $-10^{12}m_p^{-4}$, $-10^{13}m_p^{-4}$, and $-10^{14}m_p^{-4}$. From the plot we can see that, all three cases can also have inflation for $N$ up to $60$, actually very close to each other, $c_T^2$ also deviates from unity as $|M|$ grows. The perturbation variables such as the scalar power spectrum $P_S$, its spectral index $n_S$ and the tensor/scalar ratio $r$ are plotted as well, in terms of $\lambda$ and $M$, with their dependence on $\lambda$ and $M$ consistent with the above analytical analysis.

\begin{figure}
\centering
\includegraphics[scale=0.3]{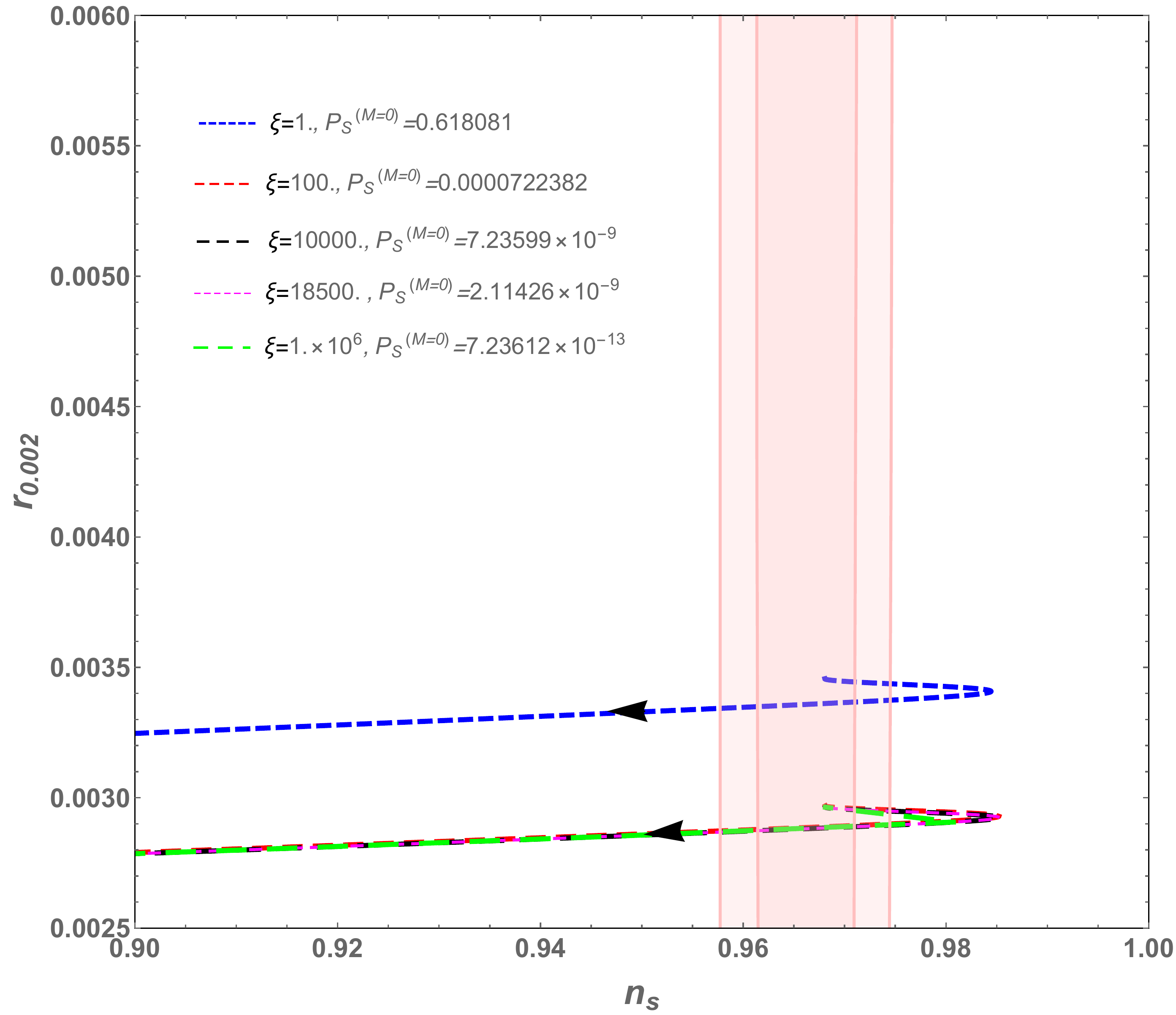}
\caption{$n_s$-$r$ constraints on Higgs inflation. We use Higgs potential $V(\phi)=\lambda\phi^4/4$ with $\lambda=0.13$. The black arrows represent the direction of $|M|$ growth. The pink region represents TT,TE,EE+lowE+lensing+BK15+BAO data constraints (68\% and 95\% CL) for $n_s$ and $r$ from 2018 Planck Collaboration data \cite{Akrami:2018odb}. Here we choose $N=60$, and $\xi$ is set as $1$, $100$, $10^4$, $1.85\times10^4$ and $10^6$, respectively. $P_S^{(M=0)}$ means the value of $P_S$ when $M=0$.}\label{higgs}
\end{figure}

In Fig. \ref{higgs} we also compare $n_S$ and $r$ in this model with the observational data. Similar to the previous one, as $|M|$ grows, both $n_S$ and $r$ decrease, with $n_S$ faster. This means, while in this case the Higgs model allows $n_S$ and $r$ fall in the favorable region given by the observations, the involvement of $M$ tend to bring a smaller value of the spectral index.
 
\subsubsection{monodromy inflation}
The next example is the so-called monodromy inflation, in which we consider a power-law inflation potential with its index being a fraction, such as:
\be
f(\phi)=1~,~~~V(\phi)=\lambda m_p^4\left(\frac{\phi}{m_p}\right)^{2/3}~.
\ee
Such a potential arises when we take into account of the UV effects in inflation cosmology, such as those from string theory, where the back-reaction of other moduli field are also incorporated although the inflationary potential plays a leading role \cite{McAllister:2014mpa}. This is reasonable as the field excursion goes beyond the Planckian energy scale \cite{Silverstein:2008sg,McAllister:2008hb}. Phenomenologically, such an inflation model is consistent with the Planck observational data, but with its location at the edge of the favorable region, indicating a little bit high spectral index. From the potential given above, we can get the initial values of $\phi$ and $\dot\phi$:
\be
\label{initialmonodromy}
\phi_{0}=\frac{m_{p}}{3}\sqrt{2(6N+1)}~,~~~\dot{\phi}_{0}=-\frac{2^{2/3}}{3^{5/6}}\frac{\sqrt{\lambda}m_{p}^{2}}{(6N+1)^{1/3}}~,
\ee
while the tensor spectrum, scalar spectrum and the tensor/scalar ratio are:
\bea
P_{S}&=&\frac{(1-2MX)\lambda^{2}m_{p}^{4}(6N+1)^{2/3}}{2^{7/3}\times3^{10/3}\pi^{2}X}~,\\
n_{S}&=&7-6\sqrt{1-2MX}\nonumber\\
&&-\frac{6^{5/3}X}{(6N+1)^{1/3}(1-2MX)\lambda m_{p}^{4}}~,\\
r&=&\frac{2^{11/3}\times3^{5/3}X}{(1-2MX)\lambda m_{p}^{4}(6N+1)^{1/3}}~,
\eea
and when $|2MX|\gg1$, we have:
\bea
P_{S}&\simeq&\frac{|M|\lambda^{2}m_{p}^{4}(6N+1)^{2/3}}{2^{4/3}\times3^{10/3}\pi^{2}}~,\\
n_{S}&\simeq&-\frac{2^{5/3}\times3^{1/6}}{(6N+1)^{1/3}}\sqrt{|M|\lambda}m_{p}^{2}~,\\
r&\simeq&\frac{2^{8/3}\times3^{5/3}}{|M|\lambda m_{p}^{4}(6N+1)^{1/3}}~.
\eea

\begin{widetext}
\begin{figure*}[ht]
\begin{center}
\includegraphics[scale=0.55]{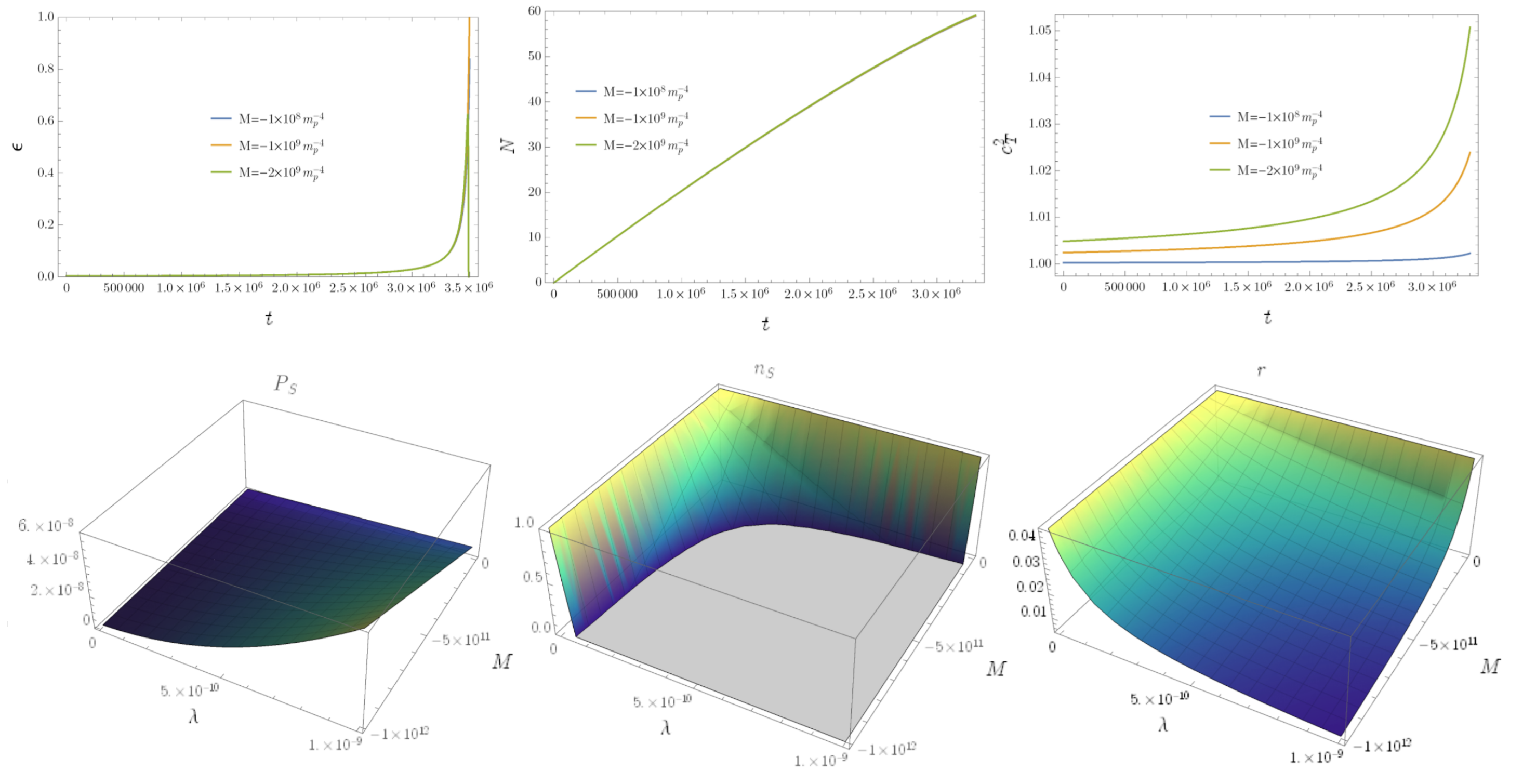}
\caption{Upper panel: The evolution of the slow-roll parameter $\epsilon$, the e-folding number $N$ as well as the sound speed squared of tensor perturbation $c_T^2$ of monodromy inflation model, with $\phi_0\simeq 8.96m_p$, $\dot\phi_0\simeq -1.56\times 10^{-6}m_p^2$, and blue, orange and green lines correspond to $M=-10^{8}m_p^{-4}$, $M=-10^{9}m_p^{-4}$, $-2\times10^{9}m_p^{-4}$, respectively. Down panel: The scalar power spectrum $P_S$, its spectral index $n_S$ and tensor/scalar ratio $r$ as functions of the parameters $\xi$ and $M$. }\label{plotmonodromy}
\end{center}
\end{figure*}
\end{widetext}

In Fig. \ref{plotmonodromy}, we plot the evolutions of slow-roll parameter $\epsilon$, the e-folding number $N$ and the sound speed squared of tensor perturbation $c_T^2$ of this model, with initial conditions chosen according to (\ref{initialmonodromy}), considering three cases of $M$ being $-10^{8}m_p^{-4}$, $-10^{9}m_p^{-4}$, and $-2\times10^{9}m_p^{-4}$. The three cases can have inflation for $N$ up to $60$ as well, and $c_T^2$ also deviates from unity as $|M|$ grows. The perturbation variables such as the scalar power spectrum $P_S$, its spectral index $n_S$ and the tensor/scalar ratio $r$ are plotted as well, in terms of $\lambda$ and $M$, with their dependence on $\lambda$ and $M$ consistent with the above analytical analysis.

\begin{figure}
\centering
\includegraphics[scale=0.3]{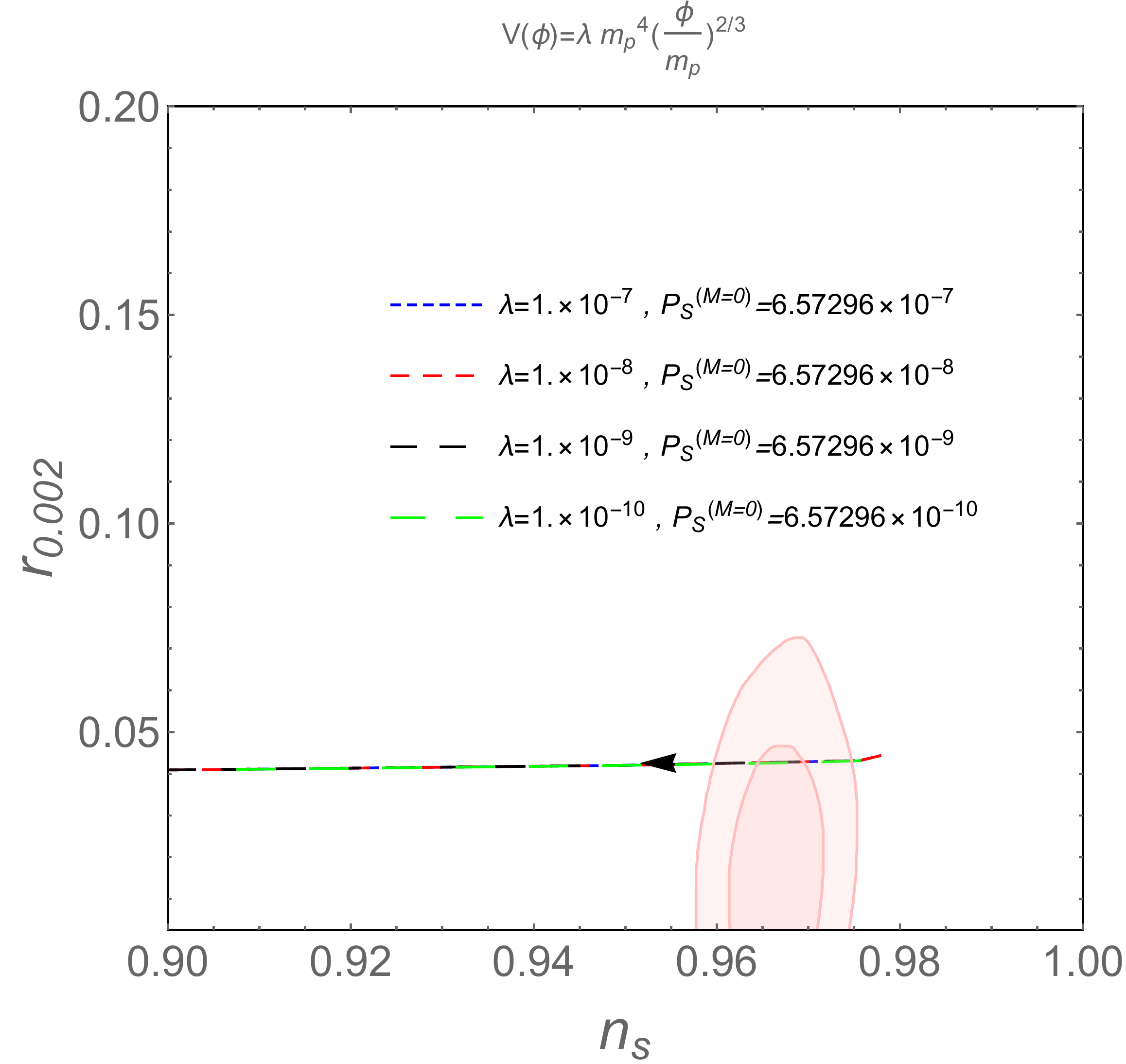}
\caption{$n_s$-$r$ constraints on monodromy inflation. The black arrows represent the direction of $|M|$ growth. The pink region represents TT,TE,EE+lowE+lensing+BK15+BAO data constraints (68\% and 95\% CL) for $n_s$ and $r$ from 2018 Planck Collaboration data \cite{Akrami:2018odb}. Here we choose $N=60$, and $\lambda$ is set as $10^{-7}$, $10^{-8}$, $10^{-9}$, and $10^{-10}$, respectively. $P_S^{(M=0)}$ means the value of $P_S$ when $M=0$.}\label{Minf}
\end{figure}

In Fig. \ref{Minf} we also compare $n_S$ and $r$ in this model with the observational data. Although the model is favorable by Planck data, note that in \cite{Akrami:2018odb} it has been shown that such a model is close to the edge of the favorable region. However, in our case, since both $n_S$ and $r$ decrease (with $n_S$ faster) as $|M|$ grows, it can make this model more close to the center of the favorable region. Namely, the model (and other models with larger $n_S$ than needed) can get improved by means of disformal coupling. 

\section{conclusion}
In this paper, we investigate the inflation models with the so-called ``disformal coupling" to gravity, which is based on the disformal transformation of two metrics. As an extension to conformal coupling, the disformal coupling also introduces the interaction of derivative terms of the field to the gravity part, which may bring interesting consequences. Therefore it is interesting to pursue for a better understanding of such a coupling.

We formulate the disformal coupling inflation model in its Jordan frame, including its Jordan frame action, the background equation of motion, as well as the perturbations, including tensor and scalar power spectrum $P_T/P_S$, scalar spectral index $n_S$ as well as tensor/scalar ratio $r$. We consider a simple example of coupling with only one additional parameter $M$, and separate our analysis into two sides.

For models where the two ``slow-roll conditions" are rigorously satisfied, we found that the apparent dependence of the disformal coupling in the expressions of $P_T$, $P_S$, $n_s$ and $r$ will be compensated by that in the expressions of initial conditions, and the final results will coincide with their Einstein frame counterpart, namely independent on the coupling. This result is in consistency with those in \cite{Minamitsuji:2014waa, Tsujikawa:2014uza, Motohashi:2015pra,Domenech:2015hka}, who claimed that the perturbations in two frames are invariant via disformal transformations. Moreover, letting $\dot\phi_0$ to be real number put a further constraint on $M$ of $0\geq M\geq -3\bar{V}\bar{K}^2/\bar{V}_\phi^2$. However, if the second ``slow-roll condition" is not applied, e.g. when we choose ``fixed values" of initial conditions independent of the disformal coupling, then the dependence of $P_S$, $n_s$ and $r$ on the coupling will appear. We exemplified this with three common models, namely chaotic inflation, Higgs inflation, and monodromy inflation, and found that for all the three models, the enlargement of $|M|$ will amplify $P_S$ and reduce $n_S$ and $r$, and vice versa. As is shown in our numerical results, this effect cannot help alleviate the dilemma of chaotic inflation, however it does have effect on making the monodromy inflation more favorable to the data. Since these models all belong to large field inflation model, albeit lack of rigid proof, it is reasonable to suspect if it is a general relationship for all the large inflation inflation models. Moreover, the sound speed of tensor and scalar perturbations are affected by the disformal coupling as well. 

Although in \cite{Minamitsuji:2014waa, Tsujikawa:2014uza, Motohashi:2015pra,Domenech:2015hka} it is shown that the perturbations are invariant under disformal transformations, it is still interesting to investigate in Jordan frame, which can show us the details of the dependence of parts of those variables on the coupling, and how it is compensated in a total. We expect it help us better understand the properties of those transformations and its applications to gravity theory and cosmology. Moreover, the superluminal behavior of the propagation speeds of scalar/tensor perturbations may also become a smoking gun. Although in general the existence of the superluminal behavior does not necessarily imply a violation of causality and formation of closed time-like curves (CTC) (see avoidance of CTC in \cite{Hawking:1991nk, Babichev:2007dw, Bruneton:2006gf, Evslin:2011vh, Burrage:2011cr}), it deserves further investigation in a general way. We will postpone these studies in future works.

\begin{acknowledgments}
We thank Taishi Katsuragawa, Jun Chen, Ze Luan, Hua Chen and Zheng Fang for helpful discussions. This work was supported by the National Natural Science Foundation of China under Grants No.~11653002 and No.~11875141.

\end{acknowledgments}

\appendix
\section{Action in the Jordan Frame}
\label{app}
In the appendix, we will give some detailed formulation through which the Jordan frame action (\ref{actionH}) is obtained. From the metric transformation of (\ref{disformalmetric}) and (\ref{inversemetric}), one can first express the Christoffel symbol in Jordan frame as:
\be
\tilde{\Gamma}_{\nu\rho}^{\mu}=\Gamma_{\nu\rho}^{\mu}+\gamma_{\nu\rho}^{\mu}~,
\ee
where 
\bea
\gamma_{\nu\rho}^{\mu}&=&f_{\nu\rho}^{\mu}+\omega_{\nu\rho}^{\mu}~,\\
f_{\rho\sigma}^{\mu}&=&\frac{1}{2}\left(g^{\mu\nu}-\frac{1}{1+\epsilon u^{2}}u^{\mu}u^{\nu}\right)[\nabla_{\rho}(u_{\nu}u_{\sigma})+\nabla_{\sigma}(u_{\nu}u_{\rho})\nonumber\\
&&-\nabla_{\nu}(u_{\rho}u_{\sigma})]~,\\
\omega_{\rho\sigma}^{\mu}&=&\delta_{\rho}^{\mu}\nabla_{\sigma}\ln\Omega+\delta_{\sigma}^{\mu}\nabla_{\rho}\ln\Omega\nonumber\\
&&-\left(g^{\mu\nu}-\frac{1}{1+\epsilon u^{2}}u^{\mu}u^{\nu}\right)\left(g_{\rho\sigma}+u_{\rho}u_{\sigma}\right)\nabla_{\nu}\ln\Omega~.
\eea
According to the definition of Ricci scalar: $R=g^{\mu\nu}R_{\mu\nu}=g^{\mu\nu}(\Gamma_{\mu\nu,\alpha}^{\alpha}-\Gamma_{\mu\alpha,\nu}^{\alpha}+\Gamma_{\mu\nu}^{\alpha}\Gamma_{\alpha\beta}^{\beta}-\Gamma_{\mu\beta}^{\alpha}\Gamma_{\nu\alpha}^{\beta})$, and after tedious calculation, we obtain the Ricci scalar:
\begin{widetext}
\bea
\tilde{R}&=&\Omega^{-2}\Bigg\{ R-\frac{2g(\phi)^{2}}{1-2g(\phi)^{2}X}R_{\mu\nu}\nabla^{\mu}\phi\nabla^{\nu}\phi+\frac{g(\phi)^{2}}{1-2g(\phi)^{2}X}[(\Box\phi)^{2}-(\nabla_{\mu}\nabla_{\nu}\phi)(\nabla^{\mu}\nabla^{\nu}\phi)]~\nonumber\\
	&&+\frac{2g(\phi)^{4}}{(1-2g(\phi)^{2}X)^{2}}[(\nabla_{\mu}X)(\nabla^{\mu}X)+(\nabla_{\mu}\phi)(\nabla^{\mu}X)\Box\phi]+\frac{8A_{\phi}(\phi)B(\phi)X-A(\phi)(3A_{\phi}(\phi)+2B_{\phi}(\phi)X)}{A(\phi)^{2}(1-2g(\phi)^{2}X)^{2}}\Box\phi~\nonumber\\
	&&-\frac{4A_{\phi}(\phi)B(\phi)-A(\phi)B_{\phi}(\phi)}{A(\phi)^{2}(1-2g(\phi)^{2}X)^{2}}(\nabla_{\mu}\phi)(\nabla^{\mu}X)-\frac{3A_{\phi}(\phi)X(A_{\phi}(\phi)-2B_{\phi}(\phi)X)}{A(\phi)^{2}(1-2g(\phi)^{2}X)^{2}}+\frac{6A_{\phi\phi}(\phi)X}{A(\phi)(1-2g(\phi)^{2}X)}\Bigg\}~.
\eea
\end{widetext}
Therefore, action (\ref{action}) can be transformed to the form:
\begin{widetext}
\bea
S_{J}&=&\int d^{4}x\sqrt{-g}\Bigg\{\frac{m_{p}^{2}}{2}A(\phi)\sqrt{1-2g(\phi)^{2}X}f(\phi)R-\frac{m_{p}^{2}}{2}\frac{g(\phi)^{2}A(\phi)f(\phi)}{\sqrt{1-2g(\phi)^{2}X}}[(\Box\phi)^{2}-(\nabla_{\mu}\nabla_{\nu}\phi)(\nabla^{\mu}\nabla^{\nu}\phi)]\nonumber\\
&&+\frac{m_{p}^{2}}{2}\left[2G_{\phi}X-\frac{-8A_{\phi}(\phi)B(\phi)X+A(\phi)(3A_{\phi}(\phi)+2B_{\phi}(\phi)X)}{A(1-2g(\phi)^{2}X)^{3/2}}f(\phi)+H\right]\Box\phi-m_{p}^{2}H_{\phi}X\nonumber\\
&&-\frac{3m_{p}^{2}}{2}\frac{A_{\phi}(\phi)f(\phi)X(A_{\phi}(\phi)-2B_{\phi}(\phi)X)}{A(\phi)(1-2g(\phi)^{2}X)^{3/2}}+\frac{3m_{p}^{2}A_{\phi\phi}(\phi)f(\phi)X}{\sqrt{1-2g(\phi)^{2}X}}+\omega(\phi)A(\phi)\frac{X}{\sqrt{1-2g(\phi)^{2}X}}\nonumber\\
&&-V(\phi)A(\phi)^{2}\sqrt{1-2g(\phi)^{2}X}\Bigg\}+S[g_{\mu\nu}, \psi]~,
\eea
\end{widetext}
where we defined
\bea
G(\phi,X)&=&\int\frac{2g(\phi)^{4}A(\phi)f(\phi)}{(1-2g(\phi)^{2}X)^{3/2}}dX\nonumber\\
&=&\frac{2g(\phi)^{2}A(\phi)f(\phi)}{\sqrt{1-2g(\phi)^{2}X}}~,\\
H(\phi,X)&=&\int\left[G_{\phi}+\frac{4A_{\phi}(\phi)B(\phi)-A(\phi)B_{\phi}(\phi)}{A(\phi)(1-2g(\phi)^{2}X)^{3/2}}f(\phi)\right]dX\nonumber\\
&=&\frac{3A_{\phi}(\phi)f(\phi)}{\sqrt{1-2g(\phi)^{2}X}}-\left(\frac{A_{\phi}(\phi)}{A(\phi)}+\frac{B_{\phi}(\phi)}{B(\phi)}+2\frac{f_{\phi}(\phi)}{f(\phi)}\right)\nonumber\\
&&\times A(\phi)f(\phi)\sqrt{1-2g(\phi)^{2}X}~.
\eea
Note that when $A=1$, such action can be reduced to that presented in \cite{Sakstein:2015jca}.

\end{document}